\documentclass[mathleft
]{an}
\usepackage{graphicx}
\usepackage{times}
\usepackage{natbib}
\bibpunct{(}{)}{;}{a}{}{,}
\begin{document}

\Pagespan{145}{}
\Yearpublication{2014}%
\Yearsubmission{2013}%
\Month{1}%
\Volume{999}%
\Issue{88}%

\title{Beyond the 2$^{nd}$ \textbf{\emph{Fermi}} Pulsar Catalog}

\author{X. Hou \thanks{Corresponding author:
  \email{hou@cenbg.in2p3.fr}\newline}
\and  D.A. Smith
\and T. Reposeur
\and R. Rousseau\thanks{for the Fermi LAT collaboration and the Pulsar Timing Consortium.}
}
\titlerunning{More {\it Fermi} Pulsars}
\authorrunning{Hou, Smith et al.}
\institute{
Centre d'\'Etudes Nucl\'eaires de Bordeaux Gradignan, IN2P3/CNRS, Universit\'e Bordeaux 1, 33175 Gradignan, France
}
\received{18 September 2013}
\accepted{26 August 2014}
\publonline{later}

\abstract{
Over thirteen times more gamma-ray pulsars have now been studied with the Large Area Telescope on NASA's {\it Fermi} satellite 
than the ten seen with the \textit{Compton Gamma-Ray Observatory} in the nineteen-nineties. 
The large sample is diverse, allowing better understanding both of the pulsars themselves and of their roles in various cosmic processes. 
Here we explore the prospects for even more gamma-ray pulsars as {\it Fermi} enters the 2$^{nd}$ half of its nominal ten-year mission. 
New pulsars will naturally tend to be fainter than the first ones discovered. 
Some of them will have unusual characteristics compared to the current population, which may help discriminate between models.
We illustrate a vision of the future with a sample of six pulsars discovered after the 2$^{nd}$ {\it Fermi} Pulsar Catalog was written.
}
\maketitle

\section{Introduction}
The First {\it Fermi} pulsar catalog \citep[][hereafter 1PC]{1PC} confirmed all ten gamma-ray pulsars published using data from the \textit{Compton Gamma-Ray Observatory} \citep{Thompson08}.
With one radio-quiet pulsar, one millisecond pulsar (MSP), and all the others young with high spindown power ($\dot E > 10^{34}$ erg s$^{-1}$), 
the CGRO objects gave a vital glimpse of the high-energy sky. Using three years of data, the Second {\it Fermi} pulsar catalog, \citep[][hereafter 2PC]{2PC}, more than doubled the
1PC tally, from 46 to 117 objects. 
Pulsars are by far the largest GeV source-class in the Milky Way.

\begin{figure*}
\centering
\includegraphics[width=0.75\textwidth]{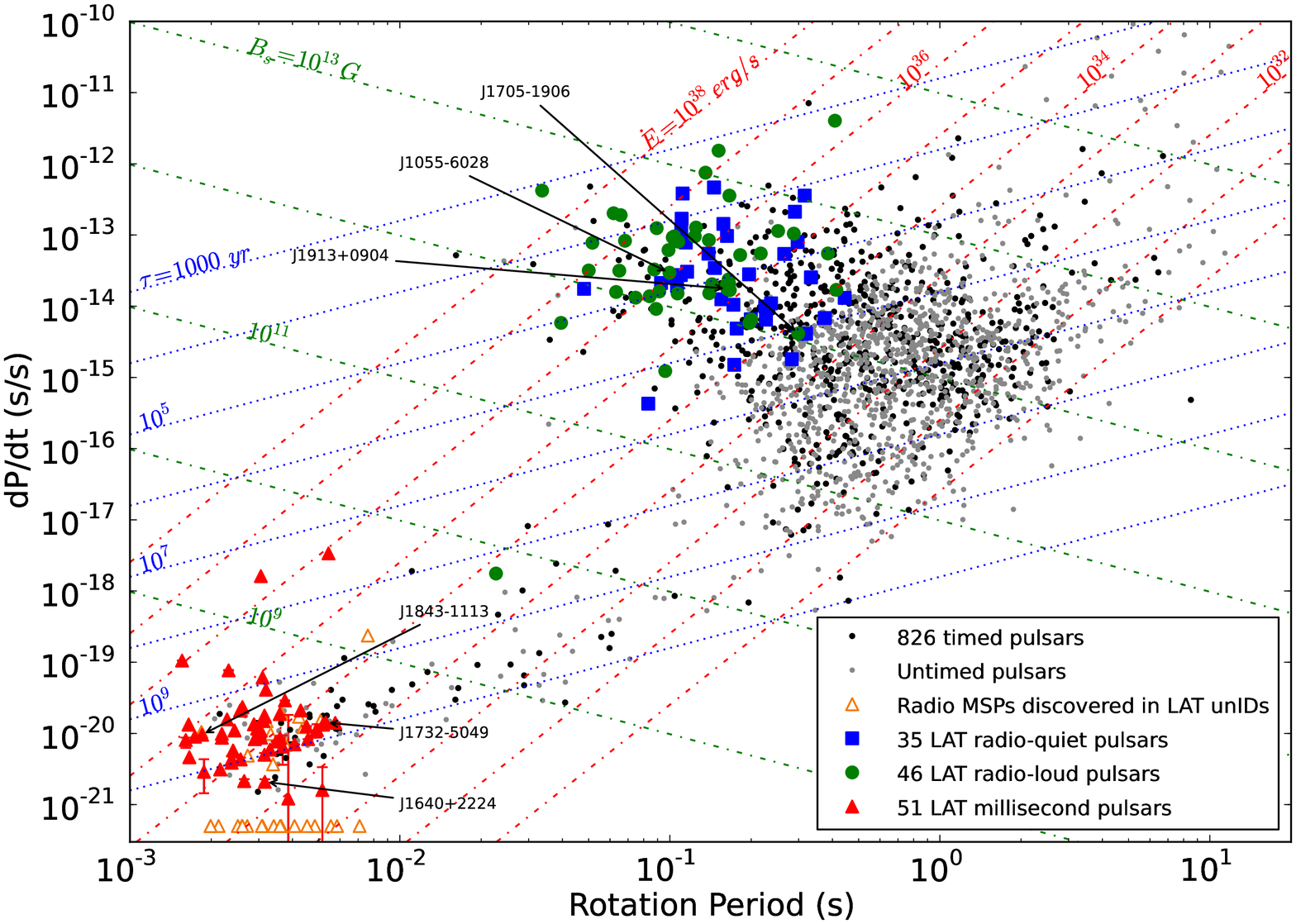}
\caption{The rate of period increase $\dot P$ vs. rotation period $P$ for 2060 pulsars,
of a total of over 2300 known (pulsars in globular clusters are not shown). Large colored markers show the 132 known gamma-ray pulsars,
highlighting six new ones discussed in this work.
``Timed'' means that we phase-folded the gamma-rays using a radio or X-ray rotation ephemeris,
but gamma pulsations were not seen. Gamma pulsations have been seen for 34 of the 54 radio MSPs 
discovered at the positions of previously unidentified LAT gamma-ray sources. 
The $\dot P$ uncertainties visible for some MSPs come from the Shklovskii correction.
New MSPs for which the spindown rate is unavailable are plotted at $\dot P = 5 \times 10^{-22}$.}
\label{PPdotplot}
\end{figure*}

As important as the large number is the large diversity, illustrated in Figure \ref{PPdotplot}. 
The CGRO pulsars pointed us to the ``guaranteed Science'' of the young radio-loud (``YRL'') pulsars that were the focus of a coordinated, large scale timing campaign \citep{TimingForFermi}. 
Less certain was whether many radio-quiet and/or MSPs would be seen. The harvest has been rich. 
Periodicity searches of the LAT (Large Area Telescope) data have yielded over 35 new young pulsars. 
Deep follow-up radio searches gave only four detections, three of which are amongst the faintest radio pulsars known (``YRQ'', for young radio-quiet).
YRL and YRQ are each about one-third of the total pulsar sample. The relative numbers, the exponentially cut-off spectral shapes, and the pulse profiles favor
gamma-ray emission from the neutron star's outer magnetosphere, as opposed to emission from near the magnetic pole region where the radio and thermal X-ray signals originate.
Several gamma-ray pulsars power GeV pulsar wind nebulae.

A surprise is that over a third of the gamma-ray pulsars are radio MSPs. Half were known before {\it Fermi}. 
But {\it Fermi} triggered a burst of radio MSP discoveries, by guiding deep searches to the positions of unidentified LAT sources. 
The current yield of 54 MSPs is a quarter of all known MSPs, outside of globular clusters.
The {\it Fermi}-triggered MSPs differ from those found in previous radio surveys. Spin periods are shorter, and
many show strong effects of the companion star's intense wind (e.g. ``black widows''). 
The latter suffer timing noise. But several LAT MSPs have stable spin periods. 
They are widespread in Galactic latitude, a benefit of the continual all-sky survey allowed by the LAT's large field-of-view and sensitivity.
This makes them particularly valuable for gravitational wave searches.
Over half of the young and recycled pulsars emit non-thermal X-rays. 

We are now half way through the nominal ten year {\it Fermi} mission. The discovery rate is slowing. 
New pulsars will mostly be less bright, making spectral and profile characterization less precise. 
Here we will argue that the new pulsars will nevertheless be valuable, 
because they will probe under-sampled parts of parameter space. 
Six new gamma-ray pulsars, half YRL and half MSPs, will serve as examples (see Table \ref{SixPsrTab}).
These ``black sheep'' may provide tests able to kill some models.

\section{Parameter Space}
A Type II supernova explosion of a massive star creates a rotating neutron star. 
To become a detectable gamma-ray pulsar, its flux must overcome the background due to
the diffuse Galactic emission and nearby sources, if and when the beam sweeps the Earth. 
The integrated energy flux is
\begin{equation}
G =  \int_{E_{min}}^{E_{max}} E {dF \over dE} = {L_\gamma \over 4\pi d^2 f_\Omega}.
\label{LumEq}
\end{equation}
The ``signature'' differential photon flux for a pulsar is ${dF \over dE} = F_0 E^\Gamma e^{-E/E_c}$, 
a hard power-law (photon index $0.5 \la \Gamma \la 2$) with an exponential cutoff. 
For weak pulsars the cutoff can be undetectable, in which case we use a simple power-law.
For bright pulsars, the observed cutoff often appears more gradual.
LAT pulsar analyses generally use $E_{min} = 100$ MeV, 
but the improved ``{\it Pass 8}'' event reconstruction soon to be released may well extend the LAT's reach to lower energies \citep{Pass8}.
$E_{max}$ is generally set to 100 or 300 GeV, a practical proxy for infinity.

$L_\gamma$ is the luminosity, and $d$ is the pulsar distance. The ``beaming factor'' $f_\Omega$, defined in 2PC,
compares the luminosity where the line-of-sight (LoS) crosses the beam with that averaged over
the full sky, for a full neutron star rotation. But $f_\Omega$ does not give information about the 
gamma-ray pulse {\it sharpness}: narrow peaks are more easily detectable than broad ones.

The background flux is strongest along the Galactic plane. 
Furthermore, multiple Coulomb scattering of the electron-positron
pairs in the LAT tracker leads to a point-spread-function (PSF) of a few degrees near
threshold energies ($\ga 100$ MeV),
decreasing to tenths of a degree at pulsar spectral cut-off energies ($0.5 \la E_c \la 5$ GeV). 
Consequently, ``hot spots'' where sources (mostly pulsars) abound, in regions such as Cygnus, have even higher background levels.
2PC Figure 16 is a sky map of the three-year pulsar sensitivity, while Figure 17 shows the longitude-averaged sensitivity range as a function of latitude, $b$.
The worst case is $G > 20 \times 10^{-12}$ erg cm$^{-2}$ s$^{-1}$ for hot spots near $b \approx 0$.
In practice, most of the known $b \approx 0$ pulsars have $G > 50 \times 10^{-12}$ erg cm$^{-2}$ s$^{-1}$. 
By $|b| = 5^\circ$, the sensitivity has improved to $10\times$ less than that.

The list of parameters affecting gamma-ray pulsar detectability thus takes form. 
An intrinsic luminosity $L_\gamma$ translates to a detectable flux $G$ depending partly on
the distance $d$ and the sky location. Peak sharpness matters. 
For $f_\Omega$, the gamma-ray pulsar ``atlases'', following the example of \citet{AtlasII}, calculate
light curves over a range of magnetic inclinations, $\alpha$, inclinations $\zeta$ to the LoS,
and accelerating gap sizes $w$, for a variety of emission models, and tabulate $f_\Omega$. 
For the majority of cases, $f_\Omega \approx 1$ and the beam factor plays a minor role in detectability.
Moving beyond 2PC, however, adding faint $f_\Omega \ll 1$ pulsars will yield a more complete sample.

Atlas light curves also show whether a given ($\alpha, \zeta, w$) configuration would give wide or sharp peaks. 
For the observed pulsar sample,
2PC provides the number of peaks, the offset $\delta$ of the gamma-ray peak from the radio pulsar, the phase separation $\Delta$ of the two gamma-ray peaks, the peak
widths, and the sharpness of the leading and trailing edges. Comparing an observed profile with model 
predictions constrains $\alpha$ and $\zeta$. Inversely, the properties of the observed gamma-ray pulsar sample can
be compared with the predictions of a population synthesis combining the neutron star distributions with a gamma-ray emission model.
Such a comparison requires that the data sample not neglect swaths of parameter space.
For example, pulsar emission can be non-pulsed, if the pulse width extends to a full rotation.
2PC includes a search for a magnetospheric signature away from the peaks, as well as a 
discussion of ``the pulsars not seen'', amongst which seem to be objects with an unpulsed pulsar-like spectrum.
T.J. Johnson et al. (in preparation, 2013) extends the atlas-like approach to the 2PC MSP sample.

Intrinsic luminosity depends on the size $w$ and shape of the region in the magnetosphere
where the electric potential is large enough to accelerate electrons to energies where they
will cascade, making even more electron-positron pairs. Curvature radiation attains higher
cutoff energies for sharper bends in the magnetic field lines -- the shape of the accelerating region
thus affects the cutoff and, inversely, if observed distributions of the spectral parameters are to be
used to discriminate between models, sample bias must be minimized.
A useful paradigm takes the open-field line voltage of an orthogonal rotating dipole to deduce
\begin{equation}
L_\gamma \propto \sqrt{\dot E}.
\label{heuristic}
\end{equation}
2PC Figure 9 plots $L_\gamma$ vs. $\dot E$, showing $\times 10$ dispersion around the above rule-of-thumb. 
The slope is not necessarily the $0.5$ of the paradigm, and the data suggest a roll-off at low $\dot E$'s.
Distance issues explain much but not all of the dispersion, and bias the $\dot E$ dependence. 
Yet, e.g. PSR J0659+1414 has a reliable parallax distance, but lies far out-of-family.
\citet{subluminous} discuss this and other pulsars where the details of the accelerating volume may
lead to significantly lower-than-typical $L_\gamma$. A pulsar data sample not biased away from these
dim cases will make model comparisons more reliable.

Striking in Figure \ref{PPdotplot} is that all gamma-ray pulsars to date lie above an apparent ``deathline''
near $\dot E \approx 3\times 10^{33}$ erg s$^{-1}$. The spindown power derives from the neutron star's magnetic field
strength $B$, period $P$, and $\alpha$, and surely this deathline indicates some minimum conditions for the gamma-ray
production mechanisms to function. Discovering new faint pulsars near the edge of death will 
probe this.

We might well also discover a few gamma-ray pulsars with $\dot E \ll  10^{33}$ erg s$^{-1}$. Outer magnetospheric processes
dominate the gamma-ray pulsar phenomenon {\it but} we also know that radio and thermal X-rays come from high energy
electrons near the neutron star magnetic poles. These electrons surely emit gamma-rays, but the intensity of those escaping the
intense magnetic field region may or may not be detectable. 
Discovering low $\dot E$ gamma-ray pulsars would mean that a much larger neutron star
population contributes to the overall gamma-ray flux in the Milky Way than now believed, with an as-yet unknown
energy spectrum. This would add yet one more complication to the search for a particle signature for Dark Matter.

\begin{table*}
\centering
\caption{Six new, radio-loud  gamma-ray faint pulsars: three are young, and three are recycled.
The separation $\Delta$ between peaks is listed only when there are $N>2$.
The $\dot E$ values with uncertainties have been Shlovskii-corrected.
The first uncertainty on $L_\gamma$ comes from the statistical error on $G$, and
the second comes from the distance error.
} 
\label{SixPsrTab}
\begin{tabular}{lrcclllcc}\hline
Pulsar name    & $P$    &$10^{-33}\dot E$&  Distance            & Gal. Lat.  & $N$ & $\delta,\,\Delta$ & $10^{12}G$                 & $10^{-33}L_\gamma$\\
               & (ms)   & (erg s$^{-1}$) &   (kpc)              & (degrees)  & (peaks) & (phase)    &(erg cm$^{-2}$ s$^{-1}$) & (erg s$^{-1}$)    \\ \hline
 J1055$-$6028  &  99.7  & 1180.          & $15.1^{+3.5}_{-6.3}$ &  -0.745    &  2 (3) & 0.13, 0.31  & $36  \pm 4   $  & $1040 \pm 120^{{+520}}_{{-670}} $\\
 J1705$-$1906  & 299.0  &  6.11          & $0.9\pm 0.1 $        &  13.026    &  1     & 0.56, --    & $ 2.0\pm 0.5 $  & $0.18 \pm 0.05 \pm 0.05$ \\
 J1913+0904    & 163.2  & 160.           & $3.0\pm 0.4 $        &  -0.684    &  2 (3) & 0.33, 0.32  & $31  \pm 5   $  & $33. \pm 3. \pm 8.$\\ 
               &        &                &                      &            &                      &  	       &	   \\
 J1640+2224    &   3.16 & $2.61 \pm 0.25$ & $1.2 \pm 0.2 $      &  38.271    &  1     & 0.48, --    & $  2.3\pm 0.6$  & $0.37 \pm 0.10 \pm 0.11 $ \\
 J1732$-$5049  &   5.31 & 3.74          & $1.4 \pm 0.2 $        &  -9.454    &  2     & 0.39, 0.27  & $  7  \pm 1  $  & $1.67 \pm 0.24 \pm 0.5 $ \\ 
 J1843$-$1113  &   1.85 & $57.8 \pm 0.02$& $1.7 \pm 0.2 $       &  -3.397    &  1     & 0.09, --    & $16  \pm 2  $  & $5.5 \pm 0.7 \pm 1.3$ \\ \hline
\end{tabular}
\end{table*}

\section{Six New Gamma-ray Pulsars}
\begin{figure*}
\centering
\includegraphics[width=0.8\textwidth]{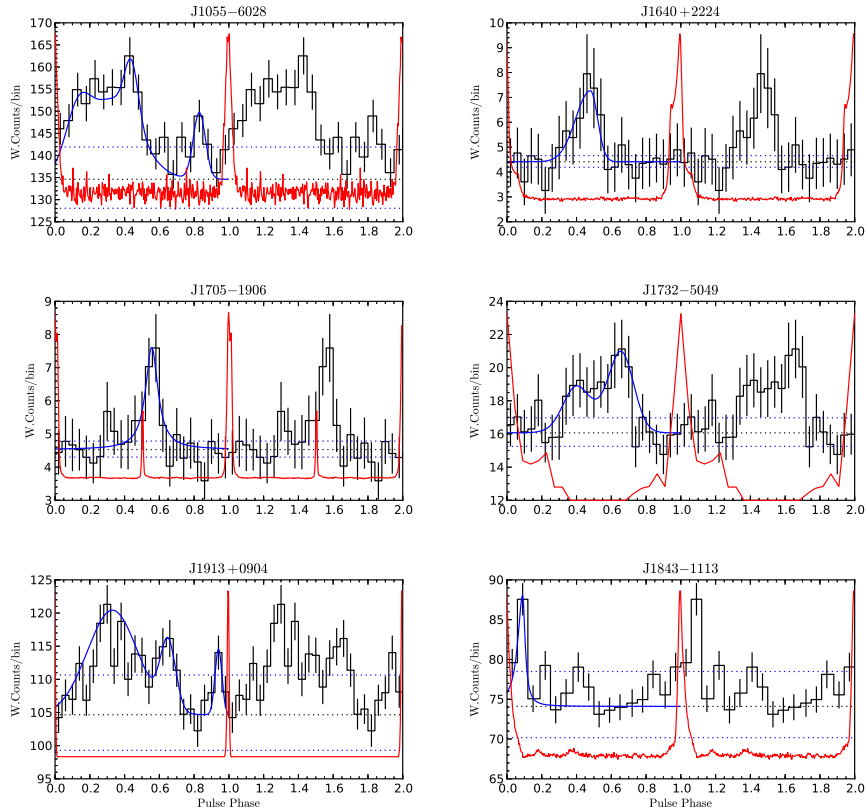}
\caption{$>100$ MeV gamma-ray pulse profiles (black histograms) for three young (left column) and 
three recycled (right column) radio-loud pulsars.
Each gamma-ray event is weighted by the probability that it came from the radio pulsar position.
The horizontal dotted lines are the background level, with $\pm 1\sigma$ uncertainties.
The blue curves are fits to the histograms, and
the red curves are the phase-aligned $\sim 1.4$ GHz radio profiles.
}
\label{lightcurves}
\end{figure*}

The weak pulsars that will dominate {\it Fermi} discoveries-to-come will cover a broader range of the parameter values discussed above,
compared to 2PC. 
Table \ref{SixPsrTab} lists a heterogeneous collection of six pulsars recently discovered to emit gamma-rays,
and Figure \ref{lightcurves} shows their lightcurves. 

We phase-folded four years of ``re-processed Pass 7'' LAT data (v15), 
requiring at least $5\sigma$ $H$-test pulsed statistical significance, as described in 2PC Section 5.
The {\em fermi} plug-in to {\it tempo2} calculates the phases \citep{gfc+12}.
The radio rotation ephemerides were provided by Parkes for J1055$-$6028, J1705$-$1906 and J1732$-$5049 \citep{Parkes} ; 
by Jodrell Bank for J1913+0904 \citep{Jodrell} ; and
by Nan\c cay for J1640+2224 and J1843$-$1113 \citep{Cognard2011}.

For the spectra, we used the standard {\it gtlike} likelihood analysis, in a circular region-of-interest of $10^\circ$ radius. 
For PSR J1640+2224, far from the bright Galactic plane, the region is centered on the pulsar. 
For the others, the circle is offset as much as $6^\circ$ (in the case of J1055$-$6028) to minimize
the influence of the plane on the fit.
An upcoming publication will detail the analysis and spectral results.
The {\it gtsrcprob} tool convolutes the $>100$ MeV spectra of the pulsar and the neighboring sources with the LAT's energy-dependent PSF 
to find the probability that a given gamma-ray event came from the pulsar. 
We weight the phase histogram entries in Figure \ref{lightcurves} by this probability.
We fit a range of profile shapes to the lightcurve, retaining the best fit, shown in blue in the figures. 
We define zero phase at the radio pulse peak. 
The absolute phase alignment of the gamma-ray data with the radio ephemeris then allows us to determine
the offset $\delta$ between the radio and gamma peaks. 

\section{A Dim Future is Bright}
Na\" ively, $P, B$, and $\alpha$ would determine the gamma-ray emission region, and thus the spectral shape and intensity as a function of neutron star longitude and latitude,
yielding the observed profile for a given $\zeta$. How well the pulsar is then seen, or not, then depends on distance $d$, the local background level, and peak sharpness. 
The first statement contains the interesting pulsar physics: the devil is in the details of the $B$ map and how it interacts with the magnetospheric current flows.
The second statement defines the observational challenge. We now use the new pulsar sample to illustrate some examples.

\subsection{Distance and Sky Location}
Fainter pulsars are farther away, all else being the same. 
PSR J1055$-$6028 is, nominally, only the third known gamma-ray pulsar beyond 10 kpc.
We won't claim that fainter gamma-pulsars will probe the neutron star population far into the Milky Way:
the $>2000$ known radio pulsars do that better. The main advantage of looking deeper is that the larger numbers of pulsars
thus accessible should better sample the different possible types. 
In 2PC, PSR J1055$-$6028 was one of the rare undetected pulsars with $\dot E > 10^{36}$ erg s$^{-1}$.
A large distance can explain why, although its single broad peak (Figure \ref{lightcurves}) and low Galactic latitude (high background)
also make detection harder. 

The distances in Table \ref{SixPsrTab} come from the radio dispersion measures (DM) and the NE2001 model \citep{Cordes2002}. 
The uncertainties are the NE2001 distances for $(1\pm 0.2)$DM. 
(For PSR J1055$-$6028, the upper error bar corresponds to the distance beyond which NE2001 models no further electrons, saturating at less than $1.2$DM.)
In most cases, the NE2001 distance corresponds to the true distance, within uncertainties,
but for any specific pulsar the NE2001 distance can be completely wrong. 
\citet{Theureau2011} detail how ``clumps'' along the LoS to PSR J0248+6021 that are unmodeled in NE2001 lead to a seeming $10\times$ overestimate
of that pulsar's distance. We have no distance estimate for most of the radio-quiet pulsars.
However, they generally have lower $\dot E$, implying modest distances, since $\dot E \propto L_\gamma \propto Gd^2$.

Unreliable distances limit the usefulness of $L_\gamma$ vs. $\dot E$ studies. Some improvements will come.
MSP J1843$-$1113 has timing residuals of $\la 1\,\mu$s, the amplitude of its annual timing parallax at the low end of its DM distance range.
MSP J1640+2224's timing may also detect parallax if the DM distance is right.
An ongoing VLBI campaign should provide distances for some MSPs and YRL pulsars to several kpc \citep{cbv+09}. 
The detailed sky surveys at radio and infrared wavelengths made to support the Planck mission provide a database that may someday
allow a three-dimensional map of Galactic electron densities far more detailed than the NE2001 model.

Of our sample, J1913+0904 is also close to the plane, and the off-pulse region is narrow. Worse, the very bright SNR W49B is $<1^\circ$ away,
explaining why detection came after 2PC in spite of a healthy $\sqrt{\dot E}/d^2$. 
The $\dot E, b$, and $d$ values for MSP J1843$-$1113 make it surprising that it took four years to show $5\sigma$ pulsations,
especially since the gamma pulse is narrow, 
but 2PC Figure 17 shows that its $G$ value is  in fact near the LAT's sensitivity limit.

Pulsations from MSP J1732$-$5049 were seen only in 2013 simply because we had never looked! Its $\dot E$ is below the threshold of the
pre-launch Timing Consortium and we had no rotation ephemeris. Once it appeared as a steady catalog source at the pulsar location,
Parkes provided one\footnote{R.N. Manchester, private communication}, and pulsations appeared immediately. MSP J1640+2224 is simply faint, in spite of being nearby, possibly due to being
right at the $\dot E$ deathline, although the geometry issues discussed by \citet{gt+14} may also play a role.
The last, PSR J1705$-$1906, is nearby, off-plane, faint, and near the deathline. 
Fortunately, its geometry is better constrained than for most pulsars.

\subsection{Geometry}
Fewer than 5\% of known pulsars have an interpulse (a radio peak a half-rotation after the first), 
reflecting the probability for $(\alpha, \zeta)$ to allow an Earth observer to see both magnetic poles \citep{interpulses}. 
PSR J1705$-$1906 is the third gamma-ray pulsar with a radio interpulse, joining PSRs J1057$-$5226, a CGRO pulsar, and J0908$-$4913.
Shown in 2PC, PSR J0908's profile looks like J1705's except that a second gamma-peak follows the second radio peak by
the same small phase interval, $\delta = 0.1$. Having defined $\phi=0$ for PSR J1705 at the position of the radio maximum, 
we obtain $\delta = 0.56$. Inverting the ``priority'' of the two radio peaks would give a slightly smaller $\delta$ than for PSR J0908.
PSR J1057$-$5226's gamma profile looks different: in 2PC a wide on-pulse region is modeled as three overlapping peaks. 
As for J1705, no gamma-ray emission appears near the other radio pulse.

\citet{B1702IP} studied J1705$-$1906 in detail and established that both $\alpha$ and $\zeta$ are near $90^\circ$, 
while \citet{5IPpsrPoln} found roughly the same inclinations for J0908$-$4913, but $\alpha \approx 75^\circ$ and $\zeta \approx 70^\circ$ for J1057$-$5226.
Of the profiles predicted for $\alpha$, $\zeta$ near $90^\circ$ by \citet{AtlasII}, 
only the Outer Gap model allows for a single gamma peak, when the gap is small. 
They set the gap size $w$ to $\eta = L_\gamma/\dot E$, which is $\sim 0.03$ for J1705 and $\sim 0.07$ for J0908.
The Atlas predicts that for the larger gap, the second gamma peak grows to be as large as the first, and the phase offset of
the first peak relative to the main radio pulse also increases. 
For J1057's angles, a single gamma peak appears for all but the largest $w$, broadening as $w$ decreases.
2PC lists $\eta \approx 0.14$ for J1057, and shows a single, broad peak, again in qualitative agreement with the Atlas.
Finally, the Atlas points to the low end of the $0.75 < f_\Omega < 0.95$  range for J1705 and J0908, and a value
closer to 1 for J1057. If accurate, then the tabulated LAT luminosity values need nearly no correction.

The other two new YRL gamma-ray pulsars, PSRs J1055$-$6028 and J1913+0904, also have wide on-pulse regions. 
We explored profile fits to a variety of pulse shapes. 
For PSR J1055$-$6028, the small peak just before the radio pulse is significant at roughly the $3\sigma$ level, or less considering the
uncertainty on the background level. In Figure \ref{lightcurves} we fit the first peak with three gaussians: the leading and trailing ones
are narrow, with the third weaker, broader one filling in the ``bridge''. A single, broad gaussian fits the structure only slightly less well. 
The three gaussian structure better resembles Atlas predictions. PSR J1913+0904's profile similarly allows some latitude in fit choices.

Knowing pulsar geometry gives both $f_\Omega$ (improving $L_\gamma$ determination) and a prediction of the gamma peak widths (improving
detectability estimates). The latter in particular is an important input to future population syntheses.
New pulsars will fill in $(\alpha, \zeta)$ regions under-sampled in 2PC.

\subsection{Deathline}
PSR J1705$-$1906 has the lowest spindown power of any YRL pulsar. 
As pulsar statistics improve, the line of pulsars near the ``empirical gamma deathline'' of $\dot E = 3\times 10^{33}$ erg s$^{-1}$ could become sharper, or it could `bleed' to lower
$\dot E$ values, depending on how strongly the gamma emission process depends on $\dot E$. 
Figure \ref{PPdotplot} is particularly rich in gray dots in this band. A specific effort from the radio pulsar timing community may be in order,
focussing on nearby, high-latitude pulsars just below the seeming $\dot E$ threshold.

Two of the new MSPs, PSRs J1640+2224 and J1732$-$5049, are right at that limit. However, the MSP deathline seems to be much lower! 
Indeed, the most recent gamma MSP discoveries are showing up at lower and lower $\dot E$ \citep{gt+14}. Exactly how low is hard to say,
because Shlovskii corrections drive it even lower, but often with very large uncertainties, and/or with an unphysical nominal value ($\dot E < 0)$. 
It is starting to look like a large majority of MSPs are gamma MSPs: the radio and gamma beams would both be very wide, and largely overlapping.
Exploring faint pulsars in the years to come will better elucidate this as well.

\section{Conclusions}
The rate for finding new gamma-ray 
pulsars\footnote{We maintain a list of published gamma-ray pulsars at \mbox{https://confluence.slac.stanford.edu/display/GLAMCOG/} -- \\ \mbox{Public+List+of+LAT-Detected+Gamma-Ray+Pulsars}} with
the {\it Fermi} LAT is higher than would be expected
from a simple $\sqrt{T}$ improvement in sensitivity as the mission livetime $T$ increases, 
because analysis breakthroughs occur. An upcoming example is that
all five years of LAT data will soon be re-processed with ``{\it Pass 8}'', with better
acceptance below 100 MeV, improving our ability to detect pulsars with low-energy spectral cutoffs.
The recent discovery of an MSP in a blind period search of gamma-ray data was possible because the orbital period
was first measured using optical measurements -- the radio detection came {\it afterwards} \citep{RayJ1311}. 
Multi-wavelength studies will continue to help us extract pulsars from the LAT data for the years to come.

Weak pulsars bring the risk of mis-characterization, for example if a second peak is too faint to be seen.
Ultimately, model predictions will incorporate detailed detector performance into comparisons with data.
Similarly, population studies will have to estimate the number of faint pulsars that are lost due to bright
neighboring sources.

Looking towards the longer term, we will have to accept, and quantify, some false detection rate in order
to obtain the largest, most complete pulsar sample possible.
To date, the LAT collaboration has insisted on $>5\sigma$ statistical pulsation significance,
preferably with a pulsar-like spectrum and pulse profile, before ``certifying'' a pulsed detection.
This was critical in the early mission, to unambiguously determine pulsar characteristics, as we did in our 1PC and 2PC catalogs.
Now we can loosen the detection criteria. This will effectively lower the minimum flux sensitivity, towards the
left in the pulsar logN-log$G$ distribution, increasing the numbers that can be seen. 
If polar cap gamma-ray emission is frequent but faint, then logN-log$G$ may increase abruptly below some flux,
allowing a burst of discoveries.
Careful Monte Carlo simulations can calculate the false detection rate, that is, the fraction of spurious pulsar-candidates.
Such contamination is a reasonable price to pay for a complete census of the Galactic neutron star population.
 
\acknowledgements
This work made extensive use of the ATNF Pulsar Catalogue\footnote{http://www.atnf.csiro.au/research/pulsar/psrcat/expert.html} \citep{ATNFcatalog}.
The Parkes radio telescope is part of the Australia Telescope which is funded by the Commonwealth Government for operation as a National Facility managed by CSIRO. 
The Nan\c{c}ay Radio Observatory is operated by the Paris Observatory, associated with the French Centre National de la Recherche Scientifique (CNRS).
The Lovell Telescope is owned and operated by the University of Manchester as part of the Jodrell Bank Centre for Astrophysics with support from 
the Science and Technology Facilities Council of the United Kingdom.

The $Fermi$ LAT Collaboration acknowledges support from a number of agencies and institutes for both development and the operation of the LAT as well as scientific data analysis. 
These include NASA and DOE in the United States, CEA/Irfu and IN2P3/CNRS in France, ASI and INFN in Italy, MEXT, KEK, and JAXA in Japan, and the K.~A.~Wallenberg Foundation, 
the Swedish Research Council and the National Space Board in Sweden. Additional support from INAF in Italy and CNES in France for science analysis during the operations phase 
is also gratefully acknowledged.


\bibliographystyle{apj}
\bibliography{2ndPulsarCatalog}


\end{document}